\documentclass[prb,reprint,showpacs]{revtex4-1}

\usepackage{hyperref}  
\usepackage[utf8]{inputenc}
\usepackage{multirow}
\usepackage{tabulary}
\usepackage[T1]{fontenc}
\usepackage[english]{babel}
\usepackage{booktabs}
\usepackage{tikz}
\usepackage{titlesec}
\usepackage{threeparttable}
\usepackage{MnSymbol}		
\usepackage{color,soul}		

\newcolumntype{K}[1]{>{\centering\arraybackslash}p{#1} }

\begin{document}

\title{Investigation of indirect excitons in bulk $2H$-MoS$_2$ using transmission electron energy-loss spectroscopy}
\author{Carsten Habenicht}
\email{c.habenicht@ifw-dresden.de}
\author{Roman Schuster}
\author{Martin Knupfer}
\author{Bernd Büchner}
\affiliation{IFW Dresden, Institute for Solid State Research,  Helmholtzstrasse 20, 01069 Dresden, Germany}
\date{\today}

\begin{abstract}

We have investigated indirect excitons in bulk $2H$-MoS$_2$ using transmission electron energy-loss spectroscopy. The electron energy-loss spectra were measured for various momentum transfer values parallel to the $\Gamma$K and $\Gamma$M directions of the Brillouin zone. The results allowed the identification of the indirect excitons between the valence band K\textsubscript{v} and conduction band $\Lambda$\textsubscript{c} points, the $\Gamma$\textsubscript{v} and K\textsubscript{c} points as well as adjacent K\textsubscript{v} and K$^{\prime}_\textrm{c}$ points. The energy-momentum dispersions for the K\textsubscript{v}-$\Lambda$\textsubscript{c}, $\Gamma$\textsubscript{v}-K\textsubscript{c} and K\textsubscript{v1}-K$^{\prime}_\textrm{c}$ excitons along the $\Gamma$K line are presented. The former two transitions exhibit a quadratic dispersion which allowed calculating their effective exciton masses based on the effective mass approximation. The K\textsubscript{v1}-K$^{\prime}_\textrm{c}$ transition follows a more linear dispersion relationship. 

\end{abstract}


\maketitle

\section{INTRODUCTION}

Layered materials such as transition metal dichalcogenides (TMDs) have been subject to a large number of scientific investigations due to their unusual electronic properties. They are used in established and experimental applications such as lubrication, catalysis, energy storage and electronics.\cite{IZYUMSKAYA_TurkishJournalofPhysics_2014__38_478, Butler_ACSnano_2013_7_4_2898, Xu_Chemicalreviews_2013_113_5_3766, Ganatra_ACSnano_2014_8_5_4074, Xu_Chemicalreviews_2013_113_5_3766} One interesting characteristics is the occurrence of excitons in semiconducting TMDs. Those electronically neutral quasi-particles are composed of electron-hole pairs in a Coulomb-bound, hydrogen-like state. They affect the optical and electronic properties of the material and their suitability for electronic applications considerably. In particular, for mono- and few-layer TMDs, they offer access to valley pseudospin which allows distinguishing excitons originating from different k-space locations\cite{Cao2012,Kim2015} and, therefore, lend themselves to potential applications in opto-electronics\cite{Butov2017SaM2} and valleytronics\cite{Xu2014}. For decades, the direct (bright) excitonic transitions in MoS$_2$ and similar TMDs have been modeled and experimentally observed using a variety of techniques.\cite{Evans_ProceedingsoftheRoyalSocietyofLondon.SeriesA.Mathematicalandphysicalsciences_1965_284_1398_402, Sobolev_OpticsandSpectroscopy_1964_18__187, Wieting_physicastatussolidi(b)_1970_37_1_353, Shepherd_JournalofPhysicsC-SolidStatePhysics_1974_7_23_4427, Yim_AppliedPhysicsLetters_2014_104_10_103114, Habenicht2015} More recently, momentum-space indirect excitons have moved in the scientific focus. \cite{Moody2016JB39, Baranowski20172M25016, Wu_Phys.Rev.B_2015_91__75310, Zhao_Nanoletters_2013_13_11_5627,Selig2016Nc13279, Zhang2017NN883} In contrast to their direct counterparts, they have an electron in a k-space location that differs from that of the associated hole. The excitation of such transitions requires not only the transfer of energy but also of an appropriate momentum. We investigated the existence of such indirect excitons in bulk $2H$-MoS$_2$ using transmission electron energy-loss spectroscopy (EELS). The results may offer benchmarks for advancing theoretical models that attempt to provide explanatory approaches that go beyond independent particle transitions and consider momentum-dependent many-particle interactions.

\begin{figure}
	\includegraphics [width=0.48\textwidth]{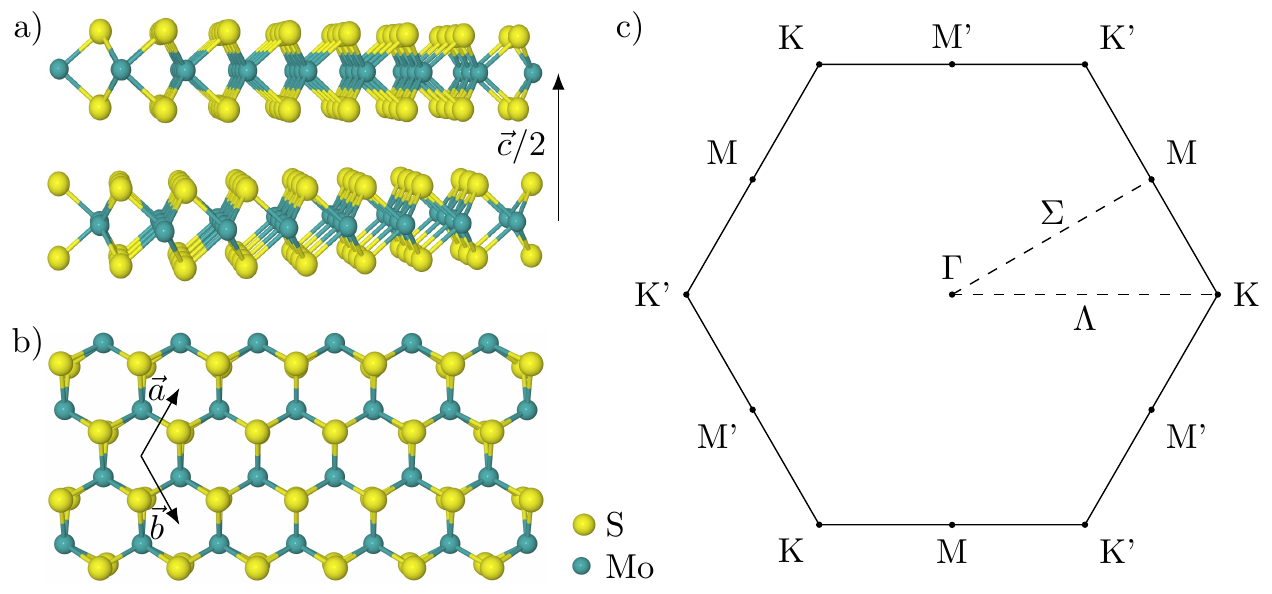}
	\caption{(Color online) Real space crystal structure of bulk $2H$-MoS$_2$ shown a) parallel and b) perpendicular to the crystal layers. c) Brillouin zone for the reduced dimensional hexagonal lattice as seen perpendicular to the in-plane direction with labels of high symmetry points and the $\Sigma$ and $\Lambda$ lines, respectively.}
	\label{fig_Crystal}
\end{figure}

The material is a semiconducting TMD forming crystals from parallel stacked layers. Each individual layer is made up of a molybdenum monolayer nested between two monolayers of sulfur bound by ionic-covalent intralayer bonds.\cite{Heda_JournalofPhysicsandChemistryofSolids_2010_71_3_187} In contrast, the layers among each other are held together by comparatively weak Van-der-Waals forces [see Fig.~\ref{fig_Crystal} a) and b) for a depiction of the crystal structure]. The properties of the material parallel to the layers vary significantly from the ones in the out-of plane direction due to the difference in the interlayer and intralayer bonding strength. The atoms in the $2H$ polytype (space group: P6$_3$/mmc, D$_{6h}^4$) exhibit a trigonal prismatic coordination within each layer with six sulfur atoms at the vertices and molybdenum at the center\cite{Dickinson_JournaloftheAmericanChemicalSociety_1923_45_6_1466} resulting in a hexagonal Brillouin zone parallel to the layers [see Fig.~\ref{fig_Crystal} c)].

\begin{figure}
	\includegraphics [width=0.48\textwidth]{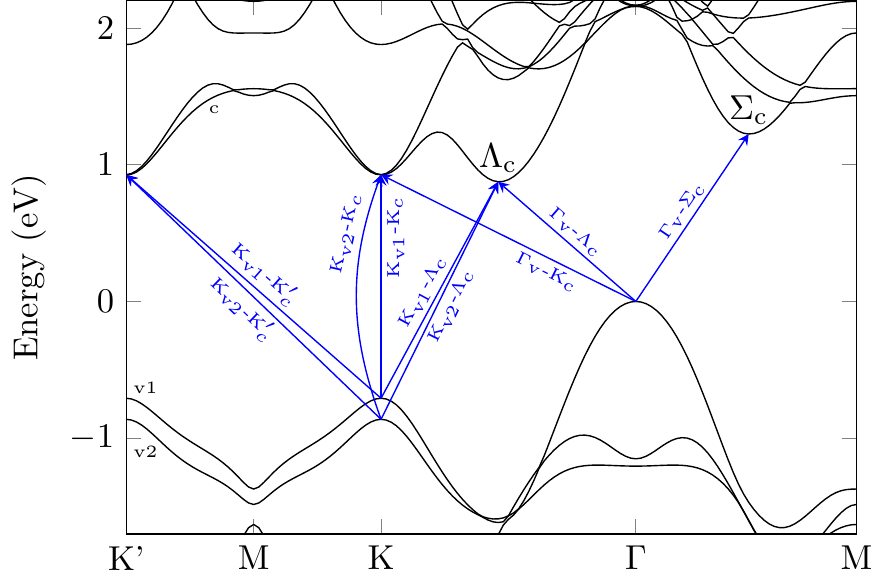}
	\caption{(Color online) Single-particle band structure for bulk $2H$-MoS$_2$ calculated using the full potential local orbital (FPLO) code\cite{Koepernik1999PRB1743} in the local-density approximation (LDA) based on a 25 x 25 x 5 point k-mesh and lattice parameters of $a=b=3.16$~Å and $c=12.32$~Å. The arrows indicate potential excitonic transitions between die valence band (indicated by subscript v) and conduction band (indicated by subscript c). The LDA computation underestimates the energy differences between the valence band maxima and the conduction band minima. Results of more involved calculation methods resulting in energy differences that are closer to experimentally observed values are presented in Table \ref{GapEnergies}.}
	\label{fig_BandStructure}
\end{figure}

Bulk $2H$-MoS$_2$ has an indirect band gap of approximately 1.2 - 1.3 eV between the $\Gamma$ point and the $\Lambda$\textsubscript{c} point of the Brillouin zone which are the absolute valence band maximum and conduction band minimum, respectively.\cite{Mattheiss_PhysicalReviewB_1973_8_8_3719, Kam_TheJournalofPhysicalChemistry_1982_86_4_463, Baglio_JournaloftheElectrochemicalSociety_1982_129_7_1461, Komsa_PhysicalReviewB_2012_86_24_241201, Cheiwchanchamnangij_PhysicalReviewB_2012_85_20_205302} See Fig.~\ref{fig_BandStructure} for a depiction of the single particle band structure. To the best of our knowledge, an exciton band structure for bulk MoS$_2$ has not been published, yet. The $\Lambda$\textsubscript{c} point is located roughly midway between the $\Gamma$ and the K point. There are also a local valence band maximum and a conduction band minimum at the K point forming a larger direct band gap.\cite{Baglio_JournaloftheElectrochemicalSociety_1982_129_7_1461, Kam_TheJournalofPhysicalChemistry_1982_86_4_463,Peelaers_PRB_2012_86_24_241401, Jiang_TheJournalofPhysicalChemistryC_2012_116_14_7664, Komsa_PhysicalReviewB_2012_86_24_241201} Another local conduction band minimum is at the $\Sigma$\textsubscript{c} point between $\Gamma$ and M. Those band extrema are favored locations for potential excitonic transitions which are indicated in Fig.~\ref{fig_BandStructure} by arrows labeled according to their transition points. The energy differences between the valence band maxima and conduction band minima published elsewhere are listed in Table~\ref{GapEnergies}. Besides the well-known direct transitions at the K and K$^{\prime}$ points, several of the indirect exciton configurations have been predicted theoretically. For monolayer MoS$_2$, Wu et al.\cite{Wu_Phys.Rev.B_2015_91__75310} suggested an exciton consisting of an electron at the K\textsubscript{c} point in the conduction band and a hole at the valence band’s $\Gamma$\textsubscript{v} point and another one between adjacent K\textsubscript{v} and K$^{\prime}_\textrm{c}$ points [The subscripts used to label Brillouin zone points refer to the valence band (v) locations of the holes and the conduction band (c) locations of the electrons]. The latter transition was also calculated by Qiu et al.\cite{Qiu2015Prl176801} Photoluminescence measurements revealed the existence of indirect excitons with energies of about 1.5 eV in bi-layer MoS$_2$. \cite{Zhao_Nanoletters_2013_13_11_5627} Comparisons of the temperature-dependent band structure calculations and photoluminescence measurements led to the conclusion that those transitions occur between the $\Gamma$\textsubscript{v} and $\Lambda$\textsubscript{c} points.\cite{Zhao_Nanoletters_2013_13_11_5627, Zhang2017NN883} Moreover, temperature- and time-dependent photoluminescence measurements on MoSe$_2$, WSe$_2$ and WS$_2$ indicate the presence of K\textsubscript{v}-K$^{\prime}_\textrm{c}$ and K\textsubscript{v}-$\Lambda$\textsubscript{c} excitons in those electronically similar materials.\cite{Selig20182M,Selig2016Nc13279, Zhang2015Prl257403}
In contrast to those techniques, which use phonons to provide the momentum transfer required to generate the spacial dislocation of the charge carriers, we were able to acquire EEL spectra of indirect excitons. This method is uniquely suited to create and measure indirect excitations as it provides not only the necessary transition energies but also allows passing a defined momentum to the electrons creating the excitons via inelastic scattering as has been shown before.\cite{Roth2015EEL37004, Schuster2018PRB41201} It permitted us for the first time to directly identify the indirect K\textsubscript{v1}-$\Lambda$\textsubscript{c}, K\textsubscript{v2}-$\Lambda$\textsubscript{c}, K\textsubscript{v1}-K$^{\prime}_\textrm{c}$, K\textsubscript{v2}-K$^{\prime}_\textrm{c}$ and $\Gamma$\textsubscript{v}-K\textsubscript{c} excitons besides the direct excitonic excitations already described in Ref. \onlinecite{Habenicht2015} by extending the investigated momentum transfer ranges.

\section{EXPERIMENT}
Natural single crystal molybdenum disulfide was acquired from Manchester Nanomaterials. Approximately 100-nm-thick films were prepared by repeatedly cleaving the bulk material along the Van-der-Waals gaps using adhesive tape. The samples were deposited on platinum transmission electron microscopy grids and transferred into the spectrometer. The instrument is a 172-keV transmission electron energy-loss spectrometer operated with an energy and momentum resolution of $\Delta E$~=~82~meV and $\Delta q$~=~0.04 Å$\textsuperscript{-1}$, respectively, under ultra-high vacuum at a temperature of 20~K achieved in a helium flow cryostat. A detailed description of the apparatus can be found in Ref. \onlinecite{Fink_AEEP_1989_75__121, Roth_JournalofElectronSpectroscopyandRelatedPhenomena_2014_195__85}. The alignment of the sample in relation to the electron beam was performed based on electron diffraction patterns. Those profiles were also used to confirm the high quality of the crystal structure. Electron energy-loss spectra between 0.2~eV and 5~eV were measured for various momentum transfer values $q$ ranging from 0.1~Å$\textsuperscript{-1}$ through 1.33~Å$\textsuperscript{-1}$ in the $\Gamma$K momentum-space direction and 1.14~Å$\textsuperscript{-1}$ in the $\Gamma$M direction. The value of 1.33~Å$\textsuperscript{-1}$ corresponds to the experimentally determined momentum space distance between the $\Gamma$ and the K point as well as between adjacent K points. 1.14~Å$\textsuperscript{-1}$ is the $\Gamma$-M separation. To detect potential beam damage of the sample, we performed measurements of diffraction patterns as well as energy-loss spectra with low momentum transfer values before and after the long-time dispersion measurements. The results showed that the pattern and spectra did not change. Consequently, the exposure to the electron beam did not appear to have an effect on the samples.

\section{RESULTS AND DISCUSSION}
\subsection{Identification of excitonic transitions}
Fig.~\ref{fig_DispBoth} shows the EELS responses measured parallel to the $\Gamma$K and $\Gamma$M directions that are the same as the $\Lambda$ and $\Sigma$ directions. It should be pointed out that a momentum transfer value for a specified direction does not represent a line segment between two particular points of the band structure. Instead, it refers to all momentum differences of the stated value in the scattering plane that are parallel to the chosen scattering direction. Consequently, a loss spectrum is the superposition of all EELS responses corresponding to a particular momentum difference and direction across the whole scattering plane which, in our case, is parallel to the planes of the real and reciprocal crystal lattice. Therefore, the spectral features cannot be assigned to specific locations within the band structure based on the momentum transfer value alone. Instead, additional criteria need to be applied. We employed a comparison of the experimental momentum transfer values and energy positions of the features in question to the calculated band structure (Fig.~\ref{fig_BandStructure}) and gap energies (Table \ref{GapEnergies}) to accomplish that task.

\subsubsection{Direct K\textsubscript{v1}-K$_c$ and K\textsubscript{v2}-K$_c$ transitions}
There are two sharp peaks at 1.98~eV and 2.16~eV of the loss spectrum for $q=0.1$~Å\textsuperscript{-1} in the $\Gamma$K direction and at 1.96~eV and 2.15~eV in the $\Gamma$M direction (Fig.~\ref{fig_DispBoth}). They represent the direct A$_1$ and B$_1$ excitonic transitions in the respective directions, located at the Brillouin zone corner points (K and K$^{\prime}$). Their peak energy differences of $\sim$180~meV arises from the splitting of the uppermost valence band at those locations. The excitations disperse to higher energies as the momentum transfer is increased until they become indiscernible at $q>0.4$~Å\textsuperscript{-1}. See Ref. \onlinecite{Habenicht2015} for a detailed analysis of those direct excitons.

\begin{table*}
\centering
\caption{Energy differences between local valence band maxima and conduction band minima in bulk $2H$-MoS$_2$ based on self-consistent GW approximation, one-shot GW (G$_0$W$_0$) approximation and the Heyd-Scuseria-Ernzerhof (HSE) functional. The numbers in parentheses are extracted from band structure plots in the respective publication.}
\label{GapEnergies}
\begin{tabular}{K{1cm}K{4.2cm}K{1.4cm}K{1.4cm}K{1.4cm}K{1.4cm}K{1.4cm}K{1.4cm}K{1.4cm}}
\toprule
\toprule
                                          		     & 				&\multicolumn{7}{c}{Energy differences (in eV)} \\\cmidrule{3-9}                                                                                                                                                                                                                                                    
 Ref.                                                  & Method                        & K\textsubscript{v1}-K$_c$ & K\textsubscript{v2}-K$_c$ & K\textsubscript{v1}-$\Lambda$\textsubscript{c} & K\textsubscript{v2}-$\Lambda$\textsubscript{c} & $\Gamma$\textsubscript{v}-K\textsubscript{c} & $\Gamma$\textsubscript{v}-$\Lambda$\textsubscript{c} & $\Gamma$\textsubscript{v}-$\Sigma$\textsubscript{c} \\
\hline 
\onlinecite{Molina-Sanchez2015SSR554}                                   & self-consistent GW      & 2.17                        & 2.41                        & 2.02                                               & (2.26)                                             & 1.59                                             & 1.44                                                     & (1.74)                                                  \\
\onlinecite{Cheiwchanchamnangij_PhysicalReviewB_2012_85_20_205302} & self-consistent GW      & 2.10                        & 2.34                        & (2.06)                                             & (2.30)                                             & (1.33)                                           & 1.29                                                     & (1.65)                                                  \\
\onlinecite{Molina-Sanchez2015SSR554}                                   & G$_0$W$_0$                    & 2.08                        & 2.32                        & 1.63                                               & (1.91)                                             & 1.69                                             & 1.24                                                     & (1.58)                                                  \\
\onlinecite{Jiang_TheJournalofPhysicalChemistryC_2012_116_14_7664} & G$_0$W$_0$                    & 2.07                        & (2.30)                      & (1.70)                                             & (1.93)                                             & (1.55)                                           & 1.23                                                     & (1.56)                                                  \\
\onlinecite{Komsa_PhysicalReviewB_2012_86_24_241201}               & HSE with spin-orbit coupling  & (2.05)                      & (2.32)                      & (2.02)                                             & (2.33)                                             & (1.47)                                           & (1.44)                                                   & (1.85)                                                  \\
\onlinecite{Hu2015SSaT55013}                                            & HSE                     & 2.01                        & (2.22)                      & 1.88                                               & (2.09)                                             & 1.43                                             & 1.30                                                     & 1.71                                                    \\
\onlinecite{Komsa_PhysicalReviewB_2012_86_24_241201}               & G$_0$W$_0$                    & (1.97)                      & (2.12)                      & (1.88)                                             & (2.02)                                             & (1.36)                                           & (1.27)                                                   & (1.58)                                                 
\\
\hline 
\bottomrule
\end{tabular}
\end{table*}

\subsubsection{Indirect K\textsubscript{v1}-$\Lambda$\textsubscript{c} and K\textsubscript{v2}-$\Lambda$\textsubscript{c} transitions}
Beginning at a momentum transfer of 0.50~Å\textsuperscript{-1}, two peaks begin to appear around 2~eV parallel to the $\Gamma$K direction. They become more pronounced at $q=0.60$~Å\textsuperscript{-1} where they reach their lowest energy position of about 1.87~eV and 2.05~eV, respectively. For larger momenta, those features disperse to higher energies until they dissipate in the continuum of other excitations at $q>0.9$~Å\textsuperscript{-1}. The momentum transfer of 0.60~Å\textsuperscript{-1}, the location of the energy dispersion minima, corresponds roughly to the $\Lambda$\textsubscript{c} point halfway between $\Gamma$ and K. Therefore, the excitations could relate to transitions from the valence band $\Gamma$ point to the $\Lambda$\textsubscript{c} point or from K to $\Lambda$\textsubscript{c}. The fact that the valence band at the K point, just like the two observed loss peaks, is split by $\sim$180~meV is an indication that the transition originates from the valence band K point and not the $\Gamma$ point. Calculated values for the single particle energy gap between the K\textsubscript{v1} and $\Lambda$\textsubscript{c} points show a relatively wide span of 430~meV from 1.63~eV through 2.06~eV (see Table~\ref{GapEnergies}). The exciton band gap of 1.87~eV (the energy position of the energetically lower peak) falls within this range and results in a possible binding energy of up to 190~meV.
For comparison, the reported ground state binding energies for the direct K\textsubscript{v1}-K$_c$ and K\textsubscript{v2}-K$_c$ excitons vary between 25~-~60~meV\cite{Yoffe_AnnualReviewofMaterialsScience_1973_3_1_147, Goto_JournalofPhysics-CondensedMatter_2000_12_30_6719, Beal_JournalofPhysicsC-SolidStatePhysics_1972_5_24_3540, Bordas_physicastatussolidi(b)_1973_60_2_505, Evans_ProceedingsoftheRoyalSocietyofLondon.SeriesA.Mathematicalandphysicalsciences_1965_284_1398_402, Cheiwchanchamnangij_PhysicalReviewB_2012_85_20_205302} and 130~-~136~meV\cite{Lee_OpticalandElectricalProperties_1976, Evans_ProceedingsoftheRoyalSocietyofLondon.SeriesA.Mathematicalandphysicalsciences_1965_284_1398_402, Evans_ProceedingsoftheRoyalSocietyofLondon.SeriesA.MathematicalandPhysicalSciences_1967_298_1452_74, Evans_physicastatussolidi(b)_1968_25_1_417}, respectively. Similarly, an exciton band gap of 2.05~eV (the energy position of the energetically higher peak) is within the calculated single particle transition energies of 1.91~eV - 2.33~eV between the K\textsubscript{v2} and $\Lambda$\textsubscript{c} points (Table~\ref{GapEnergies}). The associated binding energy is up to 280~meV. Consequently, we consider the two spectral features around 2~eV for momentum transfer values between 0.50~Å\textsuperscript{-1} and 0.9~Å\textsuperscript{-1} to be the K\textsubscript{v1}-$\Lambda$\textsubscript{c} and K\textsubscript{v2}-$\Lambda$\textsubscript{c} excitonic transitions.

\begin{figure*}
	\includegraphics [width=0.8\textwidth] {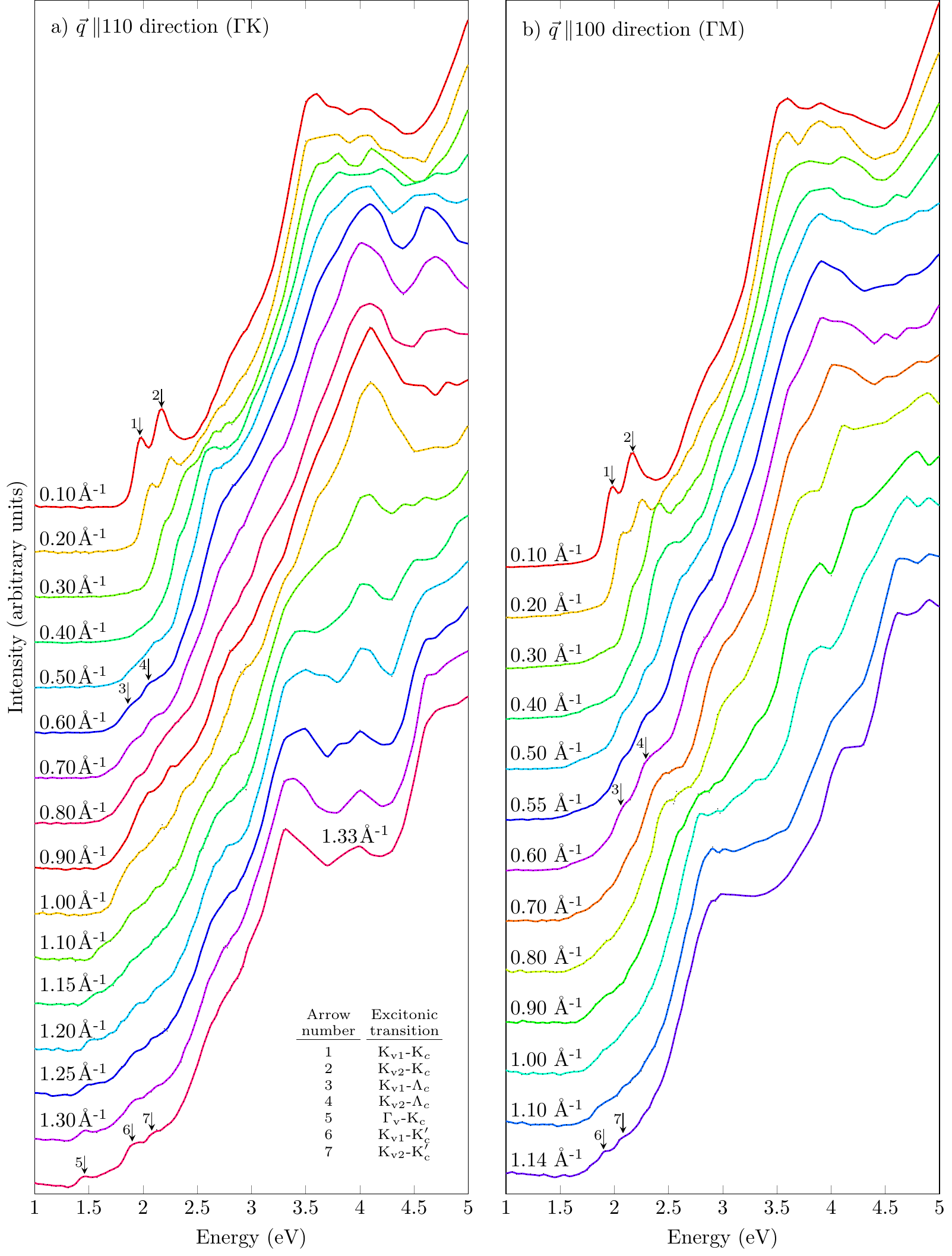}
	\caption{(Color online) Electron energy-loss spectra for bulk $2H$-MoS$_2$ measured at 20~K with $q$ parallel to the a) $\Gamma$K and b) $\Gamma$M directions for the indicated momentum transfer values. The solid spectral lines represent the binomially smoothed data (smoothing factor 1) and the black dotted lines signify the measured data points. The numbered arrows identify the energetically lowest feature associated with the respective excitonic transition (See inset legend table). For the plots, the measured intensities were normalized at 5~eV and subsequently offset for clarity.}
	\label{fig_DispBoth}
\end{figure*}

The two peaks are also visible in the $\Gamma$M direction for momentum transfer values between 0.5~Å\textsuperscript{-1} and 0.60~Å\textsuperscript{-1} [see Fig. \ref{fig_DispBoth} b)]. This suggests that the excitons have a similar spatial extension in all directions of momentum space. Because of the narrow range of momentum transfers over which the feature is visible, a dispersion is not directly noticeable in the spectra. 

\subsubsection{Indirect $\Gamma$\textsubscript{v}-K\textsubscript{c} transitions}
At a momentum transfer value of 1.15~Å\textsuperscript{-1} in the $\Gamma$K direction, a single weak peak begins to develop at 1.57~eV which disperses down to 1.46~eV at $q=1.33$~Å\textsuperscript{-1}. The latter momentum transfer value corresponds to the momentum-space difference between the $\Gamma$ and the K point. The minimum energy position and the calculated energy differences between those band structure points of 1.33~eV through 1.69~eV (Table~\ref{GapEnergies}) gives binding energies of 230~meV or less. Consequently, the feature can be assigned to the $\Gamma$\textsubscript{v}-K\textsubscript{c} excitons. A corresponding feature cannot be identified in the spectra parallel to the $\Gamma$M direction.

\subsubsection{Indirect K\textsubscript{v1}-K$^{\prime}_\textrm{c}$ and K\textsubscript{v2}-K$^{\prime}_\textrm{c}$ transitions}
Besides the peak arising from the $\Gamma$\textsubscript{v}-K\textsubscript{c} transition, two additional peaks can be identified in the vicinity of 2~eV in the spectra for $q>1.20$~Å\textsuperscript{-1}. They achieve their energy minimum of 1.90~eV and 2.08~eV at $q=1.33$~Å\textsuperscript{-1}. As before, their energy split of 180~meV indicates that the transitions originate from the valence band K point. Their lowest energy positions is below the calculated single particle energy gaps between the K points of 1.97~eV - 2.17~eV for the K\textsubscript{v1}-K$^{\prime}_\textrm{c}$ transition and 2.12~eV - 2.41~eV for the K\textsubscript{v2}-K$^{\prime}_\textrm{c}$ excitation. They are associated with potential binding energies of up to 270~meV and 330~meV for the two valence band maxima, respectively. Consequently, the measured features are classified as the indirect K\textsubscript{v1}-K$^{\prime}_\textrm{c}$ and K\textsubscript{v2}-K$^{\prime}_\textrm{c}$ excitons. The excitations are also observable in the $\Gamma$M direction indicating that their extent is comparable in all directions.
The quasi-particle band gaps of the indirect K-K$^{\prime}$ excitons in the $\Gamma$K direction are $\sim$~80~meV lower than those of the direct K point excitons. The deviation is probably due to the combined effect of changes in the interaction potential (which is inverse proportional to the momentum dependent dielectric constant) and the exchange interactions between the bound electrons and holes as their momentum space separation is larger in indirect excitons. For example, Wu et al. calculated energy differences of around 10~meV between the direct and indirect K-point excitons in monolayer MoS$_2$.\cite{Wu_Phys.Rev.B_2015_91__75310}

\begin{figure*}
	\includegraphics {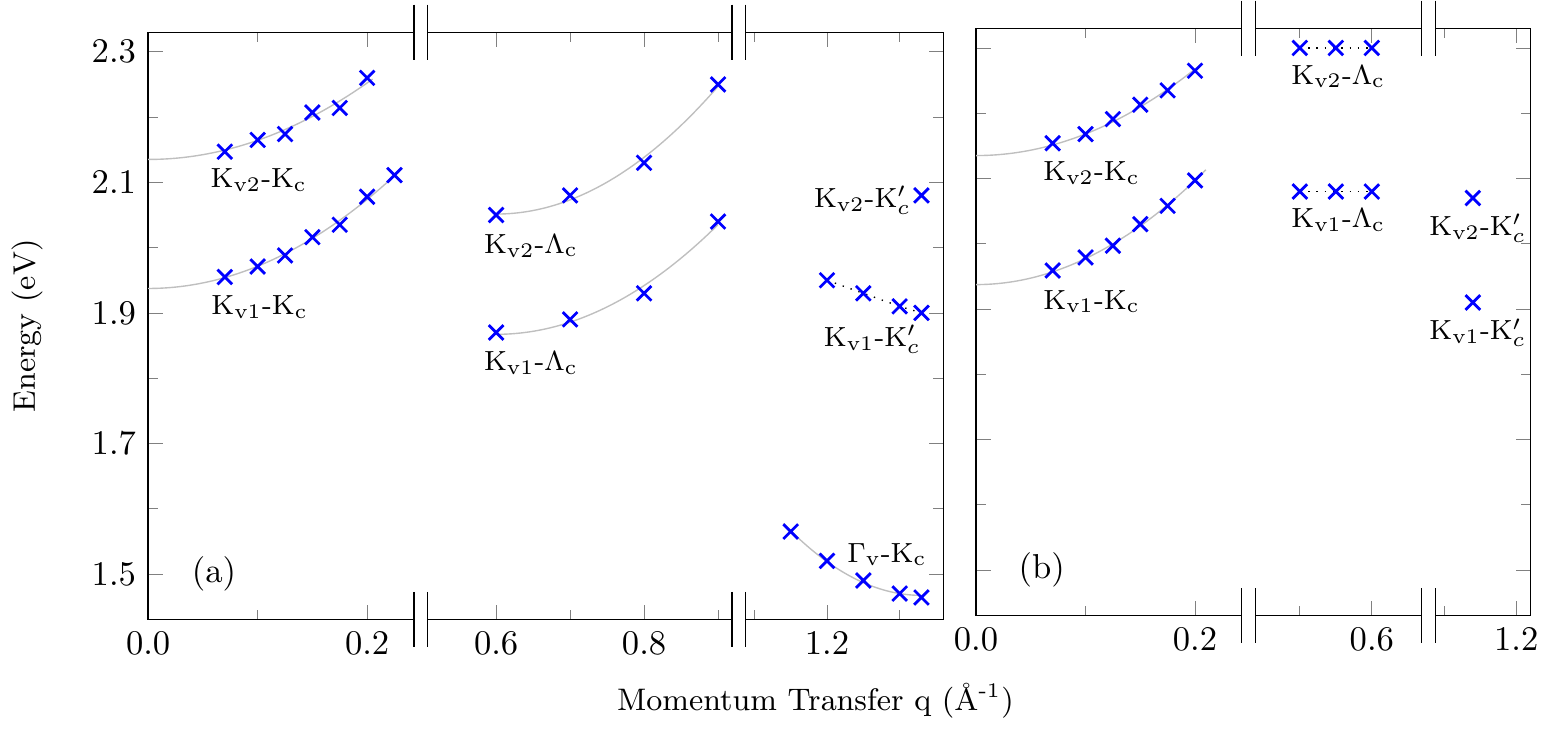}
	\caption{(Color online) Experimental dispersion of the energy-loss peak positions as function of momentum transfer (blue markers) in the (a) $\Gamma$K and (b) $\Gamma$M directions and fitted dispersion curves (solid gray lines). The peak positions were obtained by applying polynomial fits to the data EEL spectra and validated quantitatively. The dotted lines are guides for the eye only.}
	\label{fig_DispFit}
\end{figure*}

\subsubsection{Indirect $\Gamma$\textsubscript{v}-$\Lambda$\textsubscript{c} and $\Gamma$\textsubscript{v}-$\Sigma$\textsubscript{c} transitions}
A $\Gamma$\textsubscript{v}-$\Lambda$\textsubscript{c} transition should appear in the loss spectra for momentum transfer values somewhere between 0.5~Å\textsuperscript{-1} and 0.9~Å\textsuperscript{-1} at energies between 1.23~eV and 1.44~eV according to the calculation results in Table~\ref{GapEnergies}. The same should be true for the $\Gamma$\textsubscript{v}-$\Sigma$\textsubscript{c} excitation between 1.56~eV and 1.85~eV. Unfortunately, neither the spectra in the $\Gamma$K nor in the $\Gamma$M direction show any indication of such excitations. It is unclear if this is due to experimental limitations such as a low cross section or if there occur no excitons between the $\Gamma$\textsubscript{v} and $\Lambda$\textsubscript{c}/$\Sigma$\textsubscript{c}.

\subsection{Dispersions and effective masses}
To investigate the dispersion of the discussed excitons, we plotted their spectral peak energy positions as a function of the respective momentum transfer values as presented in Fig.~\ref{fig_DispFit}. Whenever the dispersions followed an approximately quadratic shape, the effective mass approximation\cite{Wang_Ultramicroscopy_1995_59_1_109} (EMA) was used to estimate the effective mass $m^*$ of the excitons from the related energy $E(q)$ and momentum values around the momentum $q_v$ associated with the energy minimum $E(q_v)$:
\begin{equation}
	E(q)=E(q_v) + \frac{\hbar^2}{2m^*}(q-q_v)^2.
	\label{equ:EMA}
\end{equation}

The fitted EMA functions are shown in Fig.~\ref{fig_DispFit} as solid gray lines and the calculation results are summarized in Table~\ref{tab:Disp}. 

\subsubsection{Direct K\textsubscript{v1}-K$_c$ and K\textsubscript{v2}-K$_c$ transitions}
The dispersion of the direct K\textsubscript{v1}-K$_c$ and K\textsubscript{v2}-K$_c$ excitons assumes roughly the form described by the EMA equation in the proximity of the K point. According to Equ. \ref{equ:EMA}, the effective masses are 1.12~m$_0$ (m$_0$ represents the electron mass) and 1.30~m$_0$ for the two excitons in the $\Gamma$K directions and 0.96~m$_0$ and 1.15~m$_0$ in the $\Gamma$M direction. An analysis of those values and a comparison of the dispersions of the indirect excitonic transitions and effective masses in bulk and monolayer material can be found in Ref. \onlinecite{Habenicht2015}.

\subsubsection{Indirect K\textsubscript{v1}-K$^{\prime}_\textrm{c}$ and K\textsubscript{v2}-K$^{\prime}_\textrm{c}$ transitions}
The K\textsubscript{v1}-K$^{\prime}_\textrm{c}$ exciton appears to display a linear dispersion along the $\Lambda$ direction. This is in contrast to the quadratic behavior of the direct K\textsubscript{v1}-K\textsubscript{c} exciton implying that the dispersion is momentum dependent. Because of the linear energy-momentum relation, the EMA may not be used to determine the effective exciton mass. Contrary to that Wu et al. calculated a quadratic dispersion for monolayer molybdenum disulfide.\cite{Wu_Phys.Rev.B_2015_91__75310} The insufficient intensity of the K\textsubscript{v2}-K$^{\prime}_\textrm{c}$ exciton peaks below $q=1.33$~Å\textsuperscript{-1} does not permit the extraction of dispersion data for that excitation.

\subsubsection{Indirect K\textsubscript{v1}-$\Lambda$\textsubscript{c} and K\textsubscript{v2}-$\Lambda$\textsubscript{c} transitions}
The K\textsubscript{v1}-$\Lambda$\textsubscript{c} and K\textsubscript{v2}-$\Lambda$\textsubscript{c} excitonic transitions exhibit a quadratic energy spread away from $q_v=0.60$~Å\textsuperscript{-1} in the $\Gamma$K direction with resulting effective masses of 2.04 m$_0$ and 1.75 m$_0$, respectively. The masses of the direct excitons are lower than those of indirect excitons. However, in contrast to the former ones, the excitation from the upper valence band has a larger effective mass than the one from the lower band. 
Because there are no published calculations of effective indirect exciton masses for bulk MoS$_2$ to which we can compare our results, we used the sum of the effective hole $m^*_h$ an electron $m^*_e$ masses as an approximation of the effective exciton mass:
\begin{equation}
	m^*\approx m^*_h + m^*_e
	\label{equ:EAHM}
\end{equation}
Utilizing HSE calculations, Peelaers and Van de Walle\cite{Peelaers_PRB_2012_86_24_241401} found effective masses of $m^*_h$~=~0.43~m$_0$ at the K point in the $\Lambda$\textsubscript{c} direction and $m^*_e$~=~0.53~m$_0$ at the $\Lambda$\textsubscript{c} point, leading to an effective exciton mass of 0.96~m$_0$ according to Equ.~\ref{equ:EAHM}. Molina-S\'anchez et al. performed G\textsubscript{0}W\textsubscript{0} calculations including spin orbit coupling (SOC) (and G\textsubscript{0}W\textsubscript{0}-SOC as well as self-consistent G\textsubscript{0}W\textsubscript{0}-SOC calculations with optimized bulk structure [optB86-vdW]) yielding an effective hole mass of 0.40~m$_0$ (and 0.39~m$_0$ as well as 0.40~m$_0$) and an effective electron mass of 0.58~m$_0$ for the same locations and directions as before.\cite{Molina-Sanchez2015SSR554} For those values, the estimated effective exciton mass was 0.98~m$_0$ (and 0.97~m$_0$ as well as 0.98~m$_0$). Yun et al.\cite{Yun_PhysicalReviewB_2012_85_3_33305} calculated direction-independent effective hole and electron masses of 0.625~m$_0$ and 0.551~m$_0$, respectively, resulting in an effective mass of 1.176~m$_0$. Those values are 33\%~-~53\% lower than our experimental results. The reason for this deviation is unclear without further momentum dependent calculations of the exciton behavior.

There is no visible dispersion of the K\textsubscript{v1}-$\Lambda$\textsubscript{c} and K\textsubscript{v2}-$\Lambda$\textsubscript{c} excitons in the spectra measured in the $\Gamma$M direction. However, between $q=0.5$~Å\textsuperscript{-1} and 0.60~Å\textsuperscript{-1} the associated features have an energy position of 2.08~eV and 2.3~eV which is 0.21~eV and 0.25~eV higher, respectively, than at the same $q$-value in the $\Gamma$K direction. That energy difference represents the energy dispersion 0.31~Å\textsuperscript{-1} away from the $\Lambda$\textsubscript{c} point along a line extending in an 75\textdegree angle from the K-$\Lambda$\textsubscript{c} line. This implies that the dispersion in this direction is stronger leading to a lower effective mass. That finding is contrary to the results of Peelaers and Van de Walle who calculated a larger in-plane effective electron mass of 0.73~m$_0$ perpendicular to the $\Gamma$K direction compared to 0.53~m$_0$ parallel to it while the hole mass at the K point is relatively isotropic.\cite{Peelaers_PRB_2012_86_24_241401}

\subsubsection{Indirect $\Gamma$\textsubscript{v}-K\textsubscript{c} transitions}
A quadratic dispersion is observable for the indirect $\Gamma$\textsubscript{v}-K\textsubscript{c} excitonic transition around $q_v=1.33$~Å\textsuperscript{-1} the (K point) parallel to the $\Lambda$ direction. Applying Equ.~\ref{equ:EMA} results in an effective mass of 1.25~m$_0$. Based on Peelaers and Van de Walle's calculations\cite{Peelaers_PRB_2012_86_24_241401}, the effective hole mass at the $\Gamma$ point and the effective electron masses at the K point along the line between those two locations are 0.62~m$_0$ and 0.47~m$_0$, respectively, adding up to an estimated effective exciton mass of 1.09~m$_0$ according to Equ.~\ref{equ:EAHM}. That value is 13\% lower than the number derived from the experimental data. Yun et al.'s\cite{Yun_PhysicalReviewB_2012_85_3_33305} calculated effective hole and electron masses of 0.711~m$_0$ and 0.821~m$_0$ result in an exciton mass of 1.532~m$_0$ which is 23\% higher than our value. Molina-S\'anchez et al. found $m^*_h$~=~0.69~m$_0$ (0.70~m$_0$) and $m^*_e$~=~0.52~m$_0$ (0.52~m$_0$) applying G\textsubscript{0}W\textsubscript{0}-SOC (and self-consistent G\textsubscript{0}W\textsubscript{0}-SOC) calculations with optB86-vdW.\cite{Molina-Sanchez2015SSR554} The associated effective exciton mass is 1.21~m$_0$ (1.22~m$_0$) deviating by only roughly 3\% from our experimental mass. The same authors also performed G\textsubscript{0}W\textsubscript{0}-SOC without optB86-vdW and computed effective hole and electron masses of 0.64~m$_0$ and 0.63~m$_0$\cite{Molina-Sanchez2015SSR554} resulting in an effective exciton mass of 1.27~m$_0$. The latter values exceeds the experimental one by only 1.6\%.

\begin{table*}
	\caption{(a) Fitting parameters for and results from the application of the effective mass approximation (Equ.~\ref{equ:EMA}) to the experimental dispersion data (see Fig.~\ref{fig_DispFit}). (b) Estimated effective exciton masses calculated according to Equ.~\ref{equ:EAHM} from the effective hole and electron masses at the respective high symmetry points. The used effective electron and hole masses are listed in Table~\ref{tab:Masses}.}
	\renewcommand{\tabcolsep}{0.15cm}
	\begin{tabular} {cccccccccc}
		\toprule
		\toprule
					&\multicolumn{4}{c}{(a) EMA fit parameters and results}	&\multicolumn{5}{c}{(b) Estimated effective exciton mass (in units of electron mass m$_0$)}\\ \cmidrule(l){2-5} \cmidrule(l){6-10}
		 			&			& \multirow{2}{*}{$\frac{\hbar^2}{2m^*}$}	&		& Effective	&\multirow{3}{*}{HSE06} &\multirow{3}{*}{FLAPW} &\multirow{3}{*}{G\textsubscript{0}W\textsubscript{0}-SOC} &\multirow{3}{*}{\parbox{1.7cm}{G\textsubscript{0}W\textsubscript{0}-SOC\\optB86bvdW}}	&\multirow{3}{*}{\parbox{2cm}{scG\textsubscript{0}W\textsubscript{0}-SOC\\optB86bvdW}} \\
     					& $E(q=0)$		&							& q$_v$	& mass $m^*$	&	&	&	&	&\\
      		Transition		& (eV) 		& (eV\,Å$^2$)					& (Å$^{-1}$)	& (m$_0$)	&	&	&	&	&\\
		\hline 
		$\Gamma$K direction:					&		&		&		&	&	&		&	&	&	\\
		K\textsubscript{v1}-K$_c$				& 1.94	& 3.41	& 0.00	& 1.12&0.90	&1.446	&1.03	&0.91	&0.92	\\
      		K\textsubscript{v2}-K$_c$				& 2.14	& 2.93 	& 0.00	& 1.30&0.90	&1.446	&1.03	&0.91	&0.92	\\
		K\textsubscript{v1}-$\Lambda$\textsubscript{c}	& 1.87	& 1.87	& 0.60	& 2.04&0.96	&1.176	&0.98	&0.97	&0.98	\\
      		K\textsubscript{v2}-$\Lambda$\textsubscript{c}	& 2.05	& 2.17	& 0.60	& 1.75&0.96	&1.176	&0.98	&0.97	&0.98	\\
		$\Gamma$\textsubscript{v}-K\textsubscript{c}	& 1.47	& 3.06	& 1.33	& 1.25&1.09	&1.532	&1.27	&1.21	&1.22	\\	
		$\Gamma$M direction:					&		&		&		&	&	&		&	&	&	\\
		K\textsubscript{v1}-K$_c$				& 1.94	& 3.99	& 0.00	& 0.96&0.91	&1.446	&1.03	&0.91	&0.92	\\
      		K\textsubscript{v2}-K$_c$				& 2.14	& 3.31 	& 0.00	& 1.15&0.91	&1.446	&1.03	&0.91	&0.92	\\
		\hline 
		\bottomrule
	\end{tabular}
	\label{tab:Disp}
\end{table*}

\begin{table*}
\begin{threeparttable}
	\caption{Calculated effective hole and electron masses (in units of electron mass m$_0$) for bulk $2H$-MoS$_2$ at the high symmetry points parallel to the directions indicated in the header subscripts. The data were obtained from the cited sources.}
	\renewcommand{\tabcolsep}{0.25cm}
	\begin{tabular} {cccccccc}
		\toprule
		\toprule
		 \multirow{2}{*}{Ref.}&\multirow{2}{*}{\parbox{4cm}{Calculation method}}	&$\Gamma$ point			&$\Lambda$\textsubscript{c} point			&\multicolumn{4}{c}{K point}\\ \cmidrule(l){5-8}
      		&						&$m^*_{h\parallel{\Gamma}K}$	&$m^*_{e\parallel\Gamma K}$	&$m^*_{h\parallel\Gamma K}$	&$m^*_{e\parallel\Gamma K}$	&$m^*_{h\parallel\Gamma M}$	&$m^*_{e\parallel\Gamma M}$\\
		\hline 
		\onlinecite{Peelaers_PRB_2012_86_24_241401}		&HSE06										&0.62		&0.53		&0.43		&0.47		&0.45\tnote{a}		&0.46\tnote{a}\\
		\onlinecite{Yun_PhysicalReviewB_2012_85_3_33305}	&FLAPW									&0.711\tnote{b}	&0.551\tnote{b}	&0.625\tnote{b}	&0.821\tnote{b}	&0.625\tnote{b}	&0.821\tnote{b}\\
      		\onlinecite{Molina-Sanchez2015SSR554}			&G\textsubscript{0}W\textsubscript{0}-SOC				&0.64		&0.58		&0.40		&0.63		&0.40		&0.63\\
		\onlinecite{Molina-Sanchez2015SSR554}			&G\textsubscript{0}W\textsubscript{0}-SOC optB86b-vdW		&0.69		&-		&0.39		&0.52		&0.39		&0.52\\
      		\onlinecite{Molina-Sanchez2015SSR554}			&Self-consistent G\textsubscript{0}W\textsubscript{0}-SOC optB86b-vdW	&0.70		&-		&0.40		&0.52		&0.40		&0.52\\
		\hline 
		\bottomrule
	\end{tabular}
	\label{tab:Masses}
\begin{tablenotes}
\item [a] Averaged effective masses for the $\Gamma$K and KM directions.
\item [b] In-plane values were reported without indicating a direction.
\end{tablenotes}
\end{threeparttable}
\end{table*}

\section{SUMMARY} 
For the first time, we have been able to directly detect indirect excitons in bulk $2H$-MoS$_2$. Electron energy-loss spectroscopy allowed identifying the K\textsubscript{v1}-$\Lambda$\textsubscript{c}, K\textsubscript{v2}-$\Lambda$\textsubscript{c}, $\Gamma$\textsubscript{v}-K\textsubscript{c}, K\textsubscript{v1}-K$^{\prime}_\textrm{c}$ and K\textsubscript{v2}-K$^{\prime}_\textrm{c}$ excitons. The data did not allow identifying the $\Gamma$\textsubscript{v}-$\Lambda$\textsubscript{c} and $\Gamma$\textsubscript{v}-$\Sigma$\textsubscript{c} excitonic transitions. We were able to determine the effective exciton masses of the K\textsubscript{v1}-$\Lambda$\textsubscript{c}, K\textsubscript{v2}-$\Lambda$\textsubscript{c} and $\Gamma$\textsubscript{v}-K\textsubscript{c} excitons of 2.04~m$_0$, 1.75~m$_0$ and 1.25~m$_0$, respectively, using the effective mass approximation. The effective masses of the two K\textsubscript{v}-$\Lambda$\textsubscript{c} transitions is higher (33\%~-~53\%) than expected values estimated by adding the effective single hole and electron masses computed for the participating band maxima and minima according to other publications. The effective mass increases for directions other than $\Lambda$. For $\Gamma$\textsubscript{v}-K\textsubscript{c}, the experimentally found effective mass agrees well with the estimated masses based on the electron and hole masses calculated by Molina-S\'anchez et al.\cite{Molina-Sanchez2015SSR554} The K\textsubscript{v1}-K$^{\prime}_\textrm{c}$ exciton assumes an approximately linear dispersion. Based on theoretical considerations, Cudazzo et al. \cite{Cudazzo2016Prl66803} determined that in single-layer materials Wannier-like excitons exhibit a parabolic dispersion while Frenkel-like excitons show a linear momentum-dependence. It would be interesting to experimentally verify if those predictions hold for bulk material. Unfortunately, our data do not allow us the determination of the spatial extent of the excitons with sufficient precision to categorize them as Wannier- or Frenkel-like. Moreover, due to our lack of access to momentum-dependent theoretical calculations, we are not able to adequately pinpoint the exciton binding energies which in theory might allow a classification. Because of those shortcomings, we are not able to associate our experimental dispersion relations for the indirect excitons with one of the two exciton types. Also, the energy gaps between the valence band maxima and conduction band minima shown in Table \ref{GapEnergies} according to the various theoretical calculation approaches exhibit deviations of up to $\sim$300 meV and therefore, may exceed typical exciton binding energies. Therefore, our data point to an incomplete understanding of the formation of exciton states and that there is a need for improving the precision of the existing theoretical models. The findings can be used as a measure to improve current band structure and exciton models. 

\begin{acknowledgments}
We thank R. H\"ubel, S. Leger, M. Naumann and U. Nitzsche for their technical assistance. R. Schuster and C. Habenicht are grateful for funding from the IFW excellence program.
\end{acknowledgments}
\newpage
%

\end{document}